%% file: z_main.tex
\documentclass{article}
\usepackage{spconfa4}
\input{definitions}

\title{HARMONICS TO THE RESCUE: WHY VOICED SPEECH IS NOT A WSS PROCESS}
\name{Giovanni Bologni$^{\sharp}$, Richard Heusdens$^{\sharp\,\flat}$ and Richard C. Hendriks$^{\sharp}$\thanks{This work was supported by NWO, the Dutch Research Council.}}
\address{$\,^{\sharp}$ Delft University of Technology, Delft, the Netherlands\\
$\,^{\flat}$ Netherlands Defence Academy, Den Helder, the Netherlands}
\begin{document}
\maketitle
%
% \begin{abstract}
% Many speech processing algorithms rely on knowledge of the statistics of the underlying speech process.
% However, despite many years of research, how to statistically model speech is still under discussion. 
% In this paper, we show that voiced speech segments can be more accurately represented as manifestations of cyclostationary (CS) processes rather than the conventional wide-sense stationary (WSS) model. 
% We provide examples of how the statistical correlation between harmonic frequencies of CS processes can be exploited to enhance system identification, a non-blind approach to estimating transfer functions, and validate our findings with both simulated and real speech data.
% \end{abstract}
\begin{abstract}
Speech processing algorithms often rely on statistical knowledge of the underlying process. 
Despite many years of research, however, the debate on the most appropriate statistical model for speech still continues.
Speech is commonly modeled as a wide-sense stationary (WSS) process.
However, the use of the WSS model for spectrally correlated processes is fundamentally wrong, as WSS implies spectral uncorrelation.
In this paper, we demonstrate that voiced speech can be more accurately represented as a cyclostationary (CS) process.
By employing the CS rather than the WSS model for processes that are inherently correlated across frequency, it is possible to improve the estimation of cross-power spectral densities (PSDs), source separation, and beamforming.
We illustrate how the correlation between harmonic frequencies of CS processes can enhance system identification, and validate our findings using both simulated and real speech data.
% Speech processing algorithms often rely on knowledge of the statistics of the underlying process, which is typically assumed to be wide-sense stationary (WSS).
% However, despite many years of research, the discussion on the most appropriate model for speech is still open
% using the WSS assumption for spectrally correlated processes is fundamentally wrong, as WSS implies spectral uncorrelatedness.
% % In this paper, we show that voiced speech segments can be more accurately represented as manifestations of cyclostationary (CS) processes rather than the conventional wide-sense stationary (WSS) model. 
% Using the more correct cyclostationary (CS) model allows to improve, for example, the estimation of cross-power spectral densities for processes that inherently have spectral correlation.
% We provide examples of how the correlation between harmonic frequencies of CS processes can be exploited to enhance system identification, a non-blind approach to estimating transfer functions, and validate our findings with both simulated and real speech data.
\end{abstract}
\vspace*{-0.1cm}
\section{Introduction}\label{sec:intro}
The complex structure of human speech poses a significant challenge for statistical modeling.
A noticeable trait of speech is non-stationarity.
To address non-stationarity, speech recordings are often divided into short segments.
Consecutive temporal frames are then treated as uncorrelated realizations of a wide-sense stationary (WSS) process \cite{kay_fundamentals_1993}. 
% Averaging these realizations helps to reduce the variance in parameter estimation.
In the frequency domain, WSS processes always decompose into distinct, asymptotically uncorrelated frequency components \cite{molisch_wireless_2011}.
Therefore, any algorithm that processes the narrowband frequency components of a signal independently, assumes, often implicitly, that the underlying process is WSS.
% However, algorithms processing narrowband frequency components independently often implicitly assume the underlying process to be WSS. 
The WSS approximation is widespread in speech-related tasks, including estimation of PSDs and transfer functions, as well as dereverberation and beamforming \cite{gannot_consolidated_2017,li_alternating_2023,moore_compact_2022}.

Because of the nearly periodic pressure waves generated by the movement of vocal folds, voiced speech segments do not behave like WSS processes.
Indeed, voiced speech is commonly represented as a combination of harmonically related sinusoidal components, known as the harmonic model \cite{mcaulay_speech_1986,krawczyk_stft_2014}.
Random signals with periodically varying first- and second-order moments are known as \emph{cyclostationary} in the wide sense and have been extensively studied, particularly in telecommunications \cite{gardner_statistical_1986,gardner_cyclostationarity_1994,gardner_cyclostationarity_2006,feher_short_1995}.
Unlike typical non-stationary models, CS models offer reliable statistical descriptors that can be computed from a single time series \cite{madisetti_cyclostationary_2009}.
Conceptually, multiple periods \emph{within} a single CS record can be thought of as multiple realizations.
Separation or detection of CS sources is achieved by leveraging their diverse periodicities, even in cases where such tasks would not be possible for WSS sources.
Another property of CS processes is that they exhibit statistical correlation over frequency.
More precisely, a signal exhibits spectral coherence if and only if it is CS \cite{gardner_cyclic_1993}.
This aligns with our understanding of voiced speech, where harmonic components at integer multiples of a fundamental frequency \emph{occur} simultaneously. 
The phenomenon of the ``missing fundamental" in pitch perception serves as a prime example, illustrating how knowledge of higher harmonics assists in inferring the frequency of an underlying fundamental periodicity \cite{moore_hearing_1995}.

Despite the recognized benefits of CS models in various fields, their application to audio processing remains largely unexplored \cite{bologni2024widebandrelativetransferfunction, dong_characterizing_2017, black_pitch_2000, jalili_speech_2018}.
\begin{figure}[tb]
\vspace{-0.2cm}
    \centering
    \hspace*{-0.2cm}
    \includegraphics[width=\linewidth]{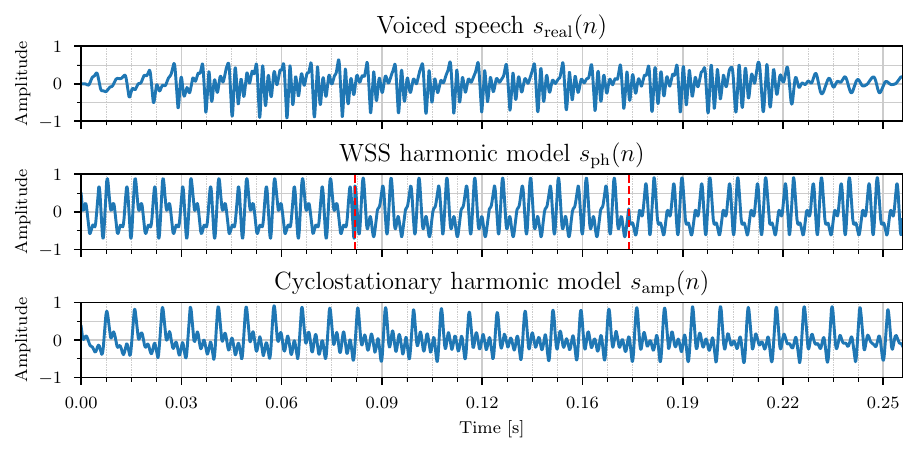}
    \caption{
    (top) voiced speech segment. 
    (mid) concatenated realizations $\sPhaseRealization$ of a WSS process; the random phase assumption introduces discontinuities at the frames' boundaries.
    (bottom) single realization $\sAmpRealization$ of a CS process.
    }
    \vspace*{-0.7cm} % reduce space between caption and following text
    \label{fig:sinusoid_challenge}
\end{figure}
This gap in research motivates our study, where we provide a theoretical justification and an experimental validation of the application of the CS model to voiced speech.
Specifically, we investigate how a CS model can capture the periodic variations inherent in voiced speech and whether leveraging this model can lead to improvements in system identification.
After outlining the theory of CS processes in \cref{sec:cyclostationarity}, we demonstrate in \cref{sec:example} how the time- and frequency-domain characteristics of CS models align with those of recorded voiced speech.
We introduce the system identification task in \cref{sec:sys_id}. 
Finally, in \cref{sec:experiments}, we verify experimentally that frequency correlation across harmonics can be exploited to improve system identification, paving the way for broader applications in audio processing tasks.
A Python implementation of all algorithms is available \cite{github_2024}.

\section{Background on Cyclostationarity}\label{sec:cyclostationarity}
We will denote random variables by capitals and the corresponding realizations by small letters.
Let $\proc{X(n), n\in \mathbb{Z}}$ be a real-valued discrete-time random process with mean $\mu_X(n) = \E{X(n)}$, and covariance
\begin{align}
\scalemath{0.93}{
r_X(n, \tau) = \E{(X(n) - \mu_X(n))(X(n + \tau) - \mu_X(n + \tau))},
}
\end{align}
for all $n, \tau \in \mathbb{Z}$.
The process is WSS if its ensemble mean $\E{X(n)} = c$ is constant over time and its autocorrelation only depends on one independent variable, \ie $r_X(n, \tau) = r_X(\tau), \forall n,\tau \in \mathbb{Z}$.
On the other hand, the process is \textit{cyclostationary} (CS) in the wide sense if both its mean and covariance function are periodic with some integer period $\funPer$:
\begin{equation}
    \mu_X(n) = \mu_X(n + \funPer), \quad
    r_X(n, \tau) = r_X(n + \funPer, \tau),
\end{equation}
for all $n, \tau \in \mathbb{Z}$.
As the mean and the covariance of a CS process are periodic in $n$ with period $\funPer$, they accept a Fourier series expansion over the set of harmonic cycles $\scalemath{0.99}{\CyclicSet = \{\alpha_{\cycidx}:2 \pi \cycidx / \funPer,~\cycidx=0,\ldots, \funPer-1\}}$.
The covariance can thus be expressed as
$\scalemathinline{0.98}{r_X(n, \tau) = \sum_{\alpha_{\cycidx} \in \CyclicSet}  \cycCorr \exp{(j \alpha_{\cycidx} n)},}$
where the Fourier coefficients, called \emph{cyclic correlations}, are given by
$\scalemathinline{0.98}{
\cycCorr = \funPer^{-1} \sum_{n=0}^{\funPer-1} r_X(n, \tau) \exp{(-j \alpha_{\cycidx} n)}.}
$
Now, suppose $\cycCorr$ is absolutely summable w.\,r.\,t.~$\tau$ for all $n$ in $\mathbb{Z}$.
By applying a Fourier transform $(\tau \to \omega)$ to $\cycCorr$, we get a function $S_X(\alpha_{\cycidx}, \omega)$ of two frequency variables, a \emph{cyclic} frequency $\alpha_{\cycidx}$ and a \emph{spectral} frequency $\omega$:
$
    S_X(\alpha_{\cycidx}, \omega) = \sum_{\tau = -\infty}^{\infty} \cycCorr \exp{(-j \omega \tau)}.
$
The quantity $S_X(\alpha_{\cycidx}, \omega)$, known as the \emph{spectral correlation density} (SCD), Loève bifrequency spectrum, or cyclic spectrum, owes its name to an alternative but equivalent definition, which can also accommodate signals with infinite energy \cite{gardner_cyclostationarity_1994}:
\begin{align}\label{eq:spec_corr_freq}
    S_X(\alpha_{\cycidx}, \omega) = \lim_{{N \to \infty}} \E{\fourier{X}_N(\omega) \fourier{X}_N^*(\omega - \alpha_{\cycidx})}, 
\end{align}
where $\fourier{X}_N(\omega) = \sum_{n=0}^{N-1} X(n) \exp{(-j\omega n)}$ is the $N$-point Fourier transform of $\scalemath{0.9}{\proc{X(n)}}$.
The SCD boils down to the conventional PSD when $\alpha_{\cycidx}=0$.
As mentioned in \cref{sec:intro}, a key property of CS processes is that they exhibit inter-frequency correlations.
In fact, $\fourier{X}_N(\omega_1)$ is correlated with $\fourier{X}_N(\omega_2)$ for $|\omega_1 - \omega_2| = \alpha_{\cycidx},~\forall\alpha_{\cycidx} \in \CyclicSet$.
By contrast, spectral components of WSS processes are asymptotically uncorrelated:
$S_X(\alpha_{\cycidx}, \omega) = 0$ and $\cycCorr = 0$ for all $\alpha_{\cycidx} \neq 0$.
Therefore, a WSS signal can be regarded as a particular CS signal for which $\cycCorr = r_X(\tau)\delta(\alpha_p)$, where $\delta(\cdot)$ is the Dirac delta.
Notice that all quantities in this section were defined for a single process $\proc{X(n)}$, but generalizing the notions to the cross-statistics between multiple processes is straightforward.
\subsection{Estimation of the spectral correlation density (SCD)}\label{ssec:est_spec_corr}
% That is, not every kind of nonstationarity has time-varying statistics that can be consistently estimated by time averages [18]. The existence of estimators of statistical functions of ACS processes that are consistent and asymptotically normal [9], [11], [29], [30] is one of the main motivations of the success of the exploitation of cyclostationarity properties in most applications. 
%
% The previous section showed that a voiced segment of speech can be modeled as a single realization of a CS process.
The definition of the SCD in \cref{eq:spec_corr_freq} involves an ensemble expectation.
Let us now introduce a method to practically estimate the cyclic spectrum of a CS process known as the \emph{time-averaged cyclic periodogram} (ACP) \cite{gardner_measurement_1986}.
%
% , which can subsequently be used for estimation, detection and system identification.
% Indeed, the practical utility of the cyclic spectrum would be limited if it could not be consistently estimated from a single record. 
% Fortunately, various methods exist for such estimation.
% The most well known is the time-averaged cyclic periodogram (ACP) estimator. 
% The time-averaged cyclic periodogram (ACP) estimator, introduced by Gardner, is the most well known method to estimate the cyclic spectrum \cite{gardner_measurement_1986}.
Essentially, the ACP replaces the expectation with a time average and coincides with Welch's PSD estimator for $\alpha_p = 0$ \cite{antoni_cyclic_2007}.
% The ACP estimator is straightforward to interpret, and its bias and variance characteristics are well understood. 
Other methods for SCD estimation
% such as the FFT accumulation method \cite{roberts_computationally_1991}, the Dirichlet method \cite{borghesani_faster_2018}, and the fast ACP \cite{alsalaet_fast_2022}, 
offer faster computations but may sacrifice interpretability \cite{roberts_computationally_1991,borghesani_faster_2018,alsalaet_fast_2022}.

% \subsubsection{Averaged cyclic periodogram}
% In \cref{eq:spec_corr_freq}, the cyclic (auto-)spectrum was defined for a single process $x(n)$;
% expanding this notion to include the cyclic (cross-)spectrum between processes $x(n)$ and $y(n)$ is straightforward.
Let $\proc{X_N(n), n \in \mathbb{Z}}$ and $\proc{Y_N(n), n \in \mathbb{Z}}$ be the finite length random processes of length $N$ sampled at sampling frequency $f_s$. 
The processes $\proc{X_N(n)}$ and $\proc{Y_N(n)}$ equal $\proc{X(n)}$ and $\proc{Y(n)}$, respectively, over the interval $\{0, \ldots, N-1\}$ and are zero otherwise.
% Consider discrete-time signals $x(n)$ and $y(n)$ of length $N$, sampled at sampling frequency $f_s$.
Processing these signals in the STFT domain, where the window length $K$ equals the DFT points and the block shift is $R$, yields a total of $L = \ceil{1 + (N - K) / R}$ frames.
To ensure ACP estimates exhibit low variance, the cyclic resolution $\Del \alpha$ must be much finer than the spectral resolution $\Del \omega$: $\Del \omega / \Del \alpha \gg 1$ \cite{gardner_measurement_1986}.
The spectral resolution is determined by the length $K$ of the DFT analysis window, $\Del \omega \approx f_s / K~[\si{\hertz}]$.
In contrast, the cyclic resolution is dictated by the total signal length, $\Del \alpha \approx {f_s}/({L R})~[\si{\hertz}]$.
This disparity in resolution levels poses challenges for implementing the frequency translation at the right-hand side of \cref{eq:spec_corr_freq}.
Therefore, it is preferable to compute the fine-grained frequency shift via time-domain modulation with cyclic frequency $\alpha_{\cycidx}$, followed by a frequency-domain transformation, thus leveraging the modulation property of the DFT: $\scalemath{0.93}{\fourier{X}(\omega - \alpha_p) \FourierPair X(n) e^{j \alpha_p n}}$.
The modulation in the time domain and its STFT counterpart are given by:
\begin{subequations}
\begin{gather}
    % y_hat[n] = y[n] * np.exp(j 2 pi alpha_hz n / fs)
    \xmod(n) = X_N(n) e^{j n \alpha_{\cycidx}}, \\
    \fourier{X}(\omega_k - \alpha_{\cycidx}, \ell) = \sum_{n=0}^{N-1} \xmod(n + \ell R){w}(n)e^{-j n \omega_k},
\end{gather}
\end{subequations}
where $\ell$ is the time-frame index and $w(n)$ represents a window function of length $K$.
% # Compute time-smoothed averaged periodogram (cyclic correlation)
% cyclic_correlation = np.zeros((x_stft.shape[0], len(alpha_vec_hz)), dtype=np.complex128)
%     for idx_aa, aa_hz in enumerate(alpha_vec_hz):
%         cyclic_correlation[:, idx_aa] = np.mean(
%             x_stft * y_stft_min_alpha[idx_aa, :].conj()
%             # * phase_correction[idx_aa]
%             , axis=1)
% By also letting $x_{k \minus \cycidx}(\ell)=x(\omega_k - \alpha_{\cycidx}, \ell)$ for conciseness, the ACP estimate is given by:
The ACP estimate is then given by:
\begin{align}\label{eq:acp_estimator}
    % \hat{S}_{yx}^{\text{acp}}(\alpha_{\cycidx}, \omega_k) = \frac{1}{L}\sum_{\ell=0}^{L-1} y_k(\ell)~\xmodk^*(\ell).
    \hat{S}_{YX}^{\text{acp}}(\alpha_{\cycidx}, \omega_k) = \frac{1}{L}\sum_{\ell=0}^{L-1} \fourier{Y}(\omega_k, \ell) \fourier{X}^*(\omega_k - \alpha_{\cycidx}, \ell).
\end{align}
\Cref{eq:acp_estimator} needs to be evaluated for all spectral bins $\omega_k = 2 \pi k / K,~k=0,\ldots,K-1$ and cyclic bins $\alpha_{\cycidx} \in \CyclicSet$.

\hide{
\subsection{Proposed sample covariance estimator}
\red{Does not work with this notation (see \cref{eq:spec_corr_time,eq:spec_corr_freq}).}
We propose an alternative method for estimating SCD that is faster but has less cyclic resolution.
Note that $k/K = f_k / f_s$.
\begin{align}
\scalemath{0.98}{
    % \hat{S}^{\text{cov}}_{yx}(k_1, k_2) = \frac{1}{L} \sum_{l=0}^{L-1} y_{k_1}(\ell) (x_{k_2}(\ell))^* e^{j2\pi\ell R (k_1 - k_2)/ K}
    \hat{S}^{\text{cov}}_{yx}(\omega_{k_1}, \omega_{k_2}) = \frac{1}{L} \sum_{\ell=0}^{L-1} y_{k_1}(\ell) (x_{k_2}(\ell))^* e^{j \ell R (\omega_{k_1} - \omega_{k_2})}
    }
\end{align}
where $k_1, k_2 = 0, \ldots, K - 1$ are the discrete frequencies, $\alpha$ is the cycle frequency, $f_s$ is the sampling frequency, $\ell=0,\ldots, L-1$ is the time-frame index, $R$ is the block shift (window length minus overlap).
}
\section{Proposed model}\label{sec:example}
% The previous sections introduced the basic theory of CS processes and outlined a method to estimate the cyclic spectrum from a sample path. 
% This section compares a WSS and a CS stochastic characterization of the harmonic model. 
In this section, we compare a WSS and a CS stochastic characterization of the harmonic model.
% We will observe that the CS candidate is more suitable for modeling voiced speech sounds.
% Let us consider two candidates for 
Let $\scalemath{0.97}{\proc{\sPhaseSingle, n \in \mathbb{Z}}}$ denote a random process 
% \begin{align}\label{eq:sin_rnd_ph_simple}
$
\sPhaseSingle = b_h \cos{(\omega_0 n h + \Phi_h)},
% s_1(n; \Phi) = \cos{(\omega_0 n + \Phi)},
% \end{align}
$
where
$b_h$ is a real amplitude,
$\omega_0 > 0$ is a normalized angular frequency,
$h \in \mathbb{N}$ is the index of the harmonic,
and $\Phi_h$ is a random variable determining the phase.
It can be shown that $\proc{\sPhaseSingle}$ is WSS when $\Phi_h$ follows, for example, a uniform distribution $\mathcal{U}(-\pi, \pi)$. 
% Given that sums of uncorrelated WSS processes are themselves WSS, a collection of random functions similar to \cref{eq:sin_rnd_ph_simple} can be combined into a new WSS process described by:
The WSS property is preserved when the uncorrelated processes corresponding to different harmonics are summed, resulting in the ``WSS harmonic model":
\begin{align}\label{eq:sin_rnd_ph_SUM}
\scalemath{0.94}{
% \check{s}_1(n; \v{\Phi}) = \sum_{h=1}^H \sPhaseSingle = \sum_{h=1}^H \cos{(\omega_0 n h + \Phi_h)},
\sPhase = \sum_{h=1}^H \sPhaseSingle = \sum_{h=1}^H b_h \cos{(\omega_0 n h + \Phi_h)}.
}
\end{align}
% where the vector $\v{\Phi} = [\Phi_1,\ldots,\Phi_H]^T$ comprises uncorrelated random variables representing the initial phases.

Alternatively, we can express each individual harmonic also as
% a random process
$
% \begin{align}\label{eq:sin_rnd_amp_simple}
% s_2(n;b(n)) = b(n) \cos{(\omega_0 n + \phi)},
\sAmpSingle = B_h(n) \cos{(\omega_0 n h+ \phi_h)},
% \end{align} 
$
where $\{B_h(n),\allowbreak n \in \mathbb{Z} \}$ denotes the amplitude characterized as a WSS process, whereas the phase $\phi_h$ remains fixed.
The mean of the process is $\mu_{\text{amp}}(n) = \mu_b \cos{(\omega_0 n h + \phi_h)}$, where $\mu_b = \E{B_h(n)}$. 
If $\mu_b = 0$, the autocovariance is given by $\autocorrAmpSingle(n, \tau) = r_b(\tau) \frac{1}{2} (\cos{(\omega_0 h \tau)} + \cos{(\omega_0 h (2n + \tau) + 2\phi_h)})$.
Notice that $\autocorrAmpSingle(n, \tau)$ \emph{cannot} be expressed as a function of a single variable $\tau$, indicating that $\proc{\sAmpSingle,\allowbreak n \in \mathbb{Z}}$ is not a WSS process.
However, the process is cyclostationary.
Specifically, if $\mu_b = 0$ then $\autocorrAmpSingle(n, \tau)$ is periodic in $n$ with period $P = 2\pi / (2\omega_0 h)$, resulting in the set of cycles $\CyclicSet = \{\pm 2\omega_0 h, 0\}$.
If $\mu_b \neq 0$, $\proc{\sAmpSingle}$ remains a CS process due to the periodicity of the mean, with the set of cycles denoted by $\CyclicSet = \{ \pm\omega_0 h, 0\}$.
% The mean of the process is $\mu_{\sAmpSingleNameOnly}(n) = \mu_b \cos{(\omega_0 n h + \phi_h)}$, 
% and its autocovariance is given by $\autocorrAmpSingle(n, \tau) = r_b(\tau) \frac{1}{2} (\cos{(\omega_0 h \tau)} + \cos{(\omega_0 h (2n + \tau) + 2\phi_h)})$.
% Notice that $\autocorrAmpSingle(n, \tau)$ \emph{cannot} be expressed as a function of a single variable $\tau$; therefore, $\sAmpSingleNameOnly$ is \emph{not} a WSS process.
% However, $\sAmpSingleNameOnly$ is a CS process.
% Indeed, if $\mu_b = 0$, the autocovariance $\autocorrAmpSingle$ exhibits periodicity with a period $P = 2\pi / (2\omega_0 h)$, with the corresponding set of cycles $\CyclicSet = \{\pm 2\omega_0 h, 0\}$.
% Alternatively, if $\mu_b \neq 0$, the set of cycles is given by $\CyclicSet = \{ \pm\omega_0 h, 0\}$ due to the periodicity of the mean.
Because sums of CS processes result in CS processes \cite[Prop.\ 1]{madisetti_cyclostationary_2009}, it is possible to form the ``CS harmonic model": 
\begin{align}\label{eq:sin_rnd_amp_SUM}
\scalemath{0.94}{
% \check{s}_2(n; \v{b}) = \sum_{h=1}^H \sAmpSingle = \sum_{h=1}^H b_h(n) \cos{(\omega_0 n h + \phi_h)},
\sAmp = \sum_{h=1}^H \sAmpSingle = \sum_{h=1}^H B_h(n) \cos{(\omega_0 n h + \phi_h)},
}
\end{align}
% where $b_h(n)$ of $\v{b}(n) = [b_1(n),\ldots,b_H(n)]^T$ represent uncorrelated WSS processes, and $\phi_1,\ldots,\phi_H$ denote deterministic quantities.
where $\proc{B_h(n)}$ represent uncorrelated WSS processes and the $\phi_h$ are deterministic.
% Representing speech as a weighted sum of sinusoidal components is commonly referred to as the ``harmonic model" \cite[Eq.\,(6)]{krawczyk_stft_2014}.
A key question arises: Among the stochastic harmonic models in \cref{eq:sin_rnd_ph_SUM,eq:sin_rnd_amp_SUM}, which of the two represents voiced speech more accurately?
In other words, among the parameters, namely the phases or the amplitudes, which exhibit randomness?

To address this, we examine an example.
The upper plot in \cref{fig:sinusoid_challenge} displays the waveform $\sReal$ of a voiced segment of speech ($f_s=\SI{48}{\kilo\hertz}$), low-pass filtered at $\SI{600}{\hertz}$ for visualization purposes. 
% of length $\SI{0.128}{\second}$
Variations in relative amplitudes among harmonics are observed, whereas the fundamental frequency appears to be constant.
The middle plot portrays three independent sample paths $\sPhaseRealization$ from \cref{eq:sin_rnd_ph_SUM}, each subjected to a rectangular window of length $K = 4096$ samples, with $\omega_0 = (2\pi / f_s) \SI{115}{\hertz}$, $H = 5$ and $b_h = 1$.
The scenario in the middle plot of \cref{fig:sinusoid_challenge} aligns with the quasi-stationarity assumption in model-based speech enhancement, where consecutive frames are regarded as independent realizations of an underlying WSS process.
Notably, at the onset of each frame, denoted by a vertical dashed red line, phase randomization introduces abrupt discontinuities, contrasting with the smooth transitions observed in the real waveform.
Lastly, the lower plot depicts a single sample path $\sAmpRealization$ from \cref{eq:sin_rnd_amp_SUM}, where each $\proc{B_h(n)}$ comprises independent Gaussian random variables distributed as $\mathcal{N}(0.5, 10)$ and filtered by a moving average process with $\floor{0.1 f_s}$ taps.
Arguably, $\sAmpRealization$ closely resembles the real waveform, suggesting that voiced speech may be more accurately represented as a single realization of a CS process rather than a collection of realizations of a WSS process.

\begin{figure}[tbh]
\vspace{-1.5em}
    \centering
    \hspace*{-0.5cm}
    \includegraphics[width=\linewidth]{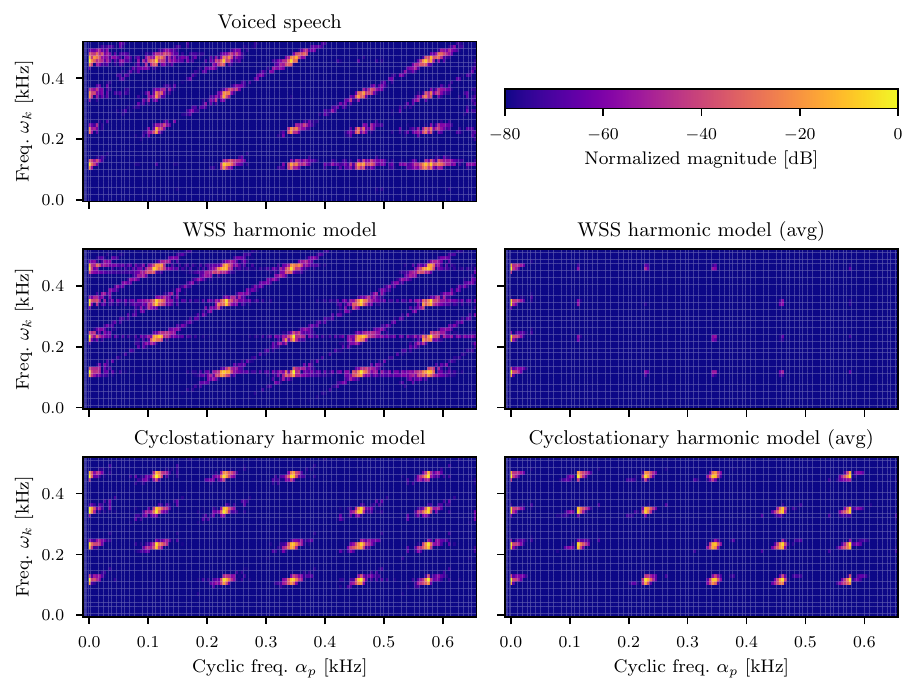}
    \caption{Magnitude of cyclic spectrum $|\hat{S}_x(\alpha_p, \omega_k)|^2$ for different signals. 
    The angular frequencies were denormalized as $f_s/{2\pi}$.
    The left column corresponds to estimates from single realization, and the right column corresponds to averages over \numRealizationsPlot realizations.}
    \label{fig:2d-scf}
    \vspace*{-1.1em} % reduce space between caption and following text
\end{figure}

The cyclic spectra of the three processes can also be analyzed to gain deeper insights.
\Cref{fig:2d-scf} illustrates the cyclic spectra magnitudes of $\sReal$, $\sPhaseRealization$, and $\sAmpRealization$ estimated by the ACP method (\cref{eq:acp_estimator}).
The cyclic spectra displayed in the left column are evaluated from a single realization of the signal with a short duration of $\approx \SI{0.25}{\second}$. 
In contrast, the plots on the right column depict estimates averaged over \numRealizationsPlot realizations to approximate the ideal spectrum.
The averaged estimate is omitted for the real signal as multiple realizations are unavailable.
The top left plot shows the cyclic spectrum of $\sReal$, denoted as $\smalleq{\hat{S}_{\text{real}}(\alpha_p, \omega_k)}$. 
The vertical slice $\smalleq{\hat{S}_{\text{real}}(0, \omega_k)}$ represents the PSD of the signal. 
The other non-zero elements of $\smalleq{\hat{S}_{\text{real}}(\alpha_p, \omega_k)}$ are found at integer multiples of the fundamental frequency $f_0 \approx \SI{115}{\hertz}$, indicating harmonic correlation.
The middle plots depict the cyclic spectrum $\smalleq{\hat{S}_{\text{ph}}(\alpha_p, \omega_k)}$ from single (left-hand) or multiple (right-hand) realizations of the WSS harmonic model $\smalleq{\proc{\sPhase}}$. 
It is worth noting that the ideal cyclic spectrum on the right exhibits nearly zero magnitudes for $\smalleq{\hat{S}_{\text{ph}}(\alpha_p \neq 0, \omega_k)}$, confirming the stationarity of the process.
Finally, the bottom plots display $\smalleq{\hat{S}_{\text{amp}}(\alpha_p, \omega_k)}$ for the CS harmonic model.
The cyclic spectra reflect the periodic nature of the signal.
Interestingly, the estimate derived from a single realization (left) closely resembles the ideal cyclic spectrum (right), confirming that a reliable descriptor of the cyclic spectrum can be obtained even from a single time frame. 

% The WSS harmonic model depicted in \cref{eq:sin_rnd_ph_SUM} facilitates mathematical analysis but inadequately represents voiced speech segments. 
The representation in \cref{eq:sin_rnd_ph_SUM} is WSS because each sinusoidal component is assumed to have a uniformly distributed phase offset \cite{hendriks_speech_2013}.
However, we argue that the apparent randomness of phase in speech is caused by STFT analysis limitations, rather than being intrinsic to speech nature.
Namely, the wrapping of the phase to its principal value between $(-\pi, \pi]$, combined with the use of fixed-duration analysis segments, irrespective of speech periodicity, results in seemingly random phase variations across frames \cite{krawczyk_stft_2014}.
% Referring phase components to a common origin, such as the beginning of the recording or voiced phoneme, alleviates the issue and reveals their correlations over frequency and time \cite{krawczyk_stft_2014}.
By contrast, the CS representation of \cref{eq:sin_rnd_amp_SUM} effectively captures the inherent periodicity of voiced speech while preserving phase relationships. 
% Furthermore, CS signals offer several advantageous properties. 
% First, the cyclic spectrum of a CS signal remains unaffected by the addition of WSS noise for all cyclic frequencies other than 0, facilitating parameter estimation (cf. \cref{sec:sys_id}). 
% Secondly, the properties of a CS signal can be consistently estimated from a single record, contrasting with WSS signals, whose properties are only asymptotically consistent.
\section{System identification}\label{sec:sys_id}
As discussed in the previous sections, a voiced segment of speech can be modeled as a CS process, and an estimate of its cyclic spectrum can be obtained via the ACP estimator.
This section shows how system identification can be improved by exploiting correlation between harmonic components in the cyclic spectrum.
% Let us now exploit the frequency correlation property of cyclostationary processes for the system identification task.
In the STFT domain, let $\fourier{S}(\omega_k, \ell) = \fourier{S}_k(\ell)$ be the clean input signal at frequency bin $k$ and frame $\ell$, $\fourier{N}_k(\ell)$ the input noise, $a_k$ the transfer function to be estimated, and $\fourier{V}_k(\ell)$ the output noise, so that
\vspace{-0.3em}
\begin{align}
    \fourier{Z}_k(\ell) = \fourier{S}_k(\ell) + \fourier{N}_k(\ell), \quad
    \fourier{X}_k(\ell) = \fourier{S}_k(\ell) \, a_k + \fourier{V}_k(\ell).
\end{align}
When the excitation signal $\fourier{S}$ and the output noise $\fourier{V}$ are WSS and uncorrelated, and the input noise $\fourier{N}$ is absent, the optimal estimate for the system transfer function in the MMSE sense is given by ratio between the input-output cross-PSD and the input PSD, known as the Wiener estimator \cite{wiener_extrapolation_1949}:
\vspace{-0.3em}
% $
\begin{align}\label{eq:wie_est}
    % \hat{a}_k = \frac{\hat{S}_{xz}(k)}{\hat{S}_{z}(k)}
    \hat{a}^{\text{Wie}}_k = {\hat{S}_{XZ}(\omega_k)}/{\hat{S}_{Z}(\omega_k)}.
\end{align}
% $
If the excitation signal is CS with cyclic frequency $\alpha_0$ while the noises are stationary, 
the cyclic spectrum of the noisy input at cyclic frequency $\alpha_0$ will not be influenced by noise, \ie $S_z(\omega_k, \alpha_0) = S_s(\omega_k, \alpha_0)$.
This observation led Gardner to the design of an estimator that relies on the cyclic spectra
% , given by
% \begin{align}
% $
    % \hat{a}^{\text{Gar}}_k = {\hat{S}_{xz}(\omega_k, \alpha_0)} / {\hat{S}_{z}(\omega_k, \alpha_0)}
% $
 \cite{gardner_identification_1990}.
Based on this, Antoni \emph{et al.~}proposed an improved system identification algorithm that combines the estimates from every integer multiple of the fundamental cyclic frequency \cite{antoni_h_alpha_2004}:
% Their method assumes that the system input exhibits cyclostationarity with a known cyclic frequency $\alpha_0$ and that the input noise is independent of the system input and does not exhibit cyclostationarity with the same cycle frequency $\alpha_0$.
% Their method assumes that $(i)$ the system input exhibits cyclostationarity with a known cyclic frequency $\alpha_0$, $(ii)$ the cyclic spectra exist and are non-zero at the frequencies of interest, and $(iii)$ the input noise has average value zero, is independent of system input, and does not exhibit cyclostationarity with the same cycle frequency $\alpha_0$ as that of the input.
% A weighted average of the estimates across harmonic components gives the transfer function estimator:
\vspace{-0.3em}
\begin{align}\label{eq:h_alpha_est}
    \hat{a}^{\text{Ant}}_k = \sum_{\alpha_{\cycidx} \in \CyclicSet} \beta_{\alpha_{\cycidx}} \frac{\hat{S}_{XZ}(\omega_k, \alpha_{\cycidx})}{\hat{S}_{Z}(\omega_k, \alpha_{\cycidx})}, \quad \sum_{\alpha_{\cycidx} \in \CyclicSet} \beta_{\alpha_{\cycidx}} = 1.
\end{align}
The optimal coefficients that minimize the variance of the estimator depend on the statistics of the clean input $\fourier{S}_k(\ell)$.
Since those statistics are unavailable, we resort to the input-output cross-statistics by defining
% \begin{align}
$
\beta_{\alpha_{\cycidx}} = \gamma^2_{\cycidx} / \sum_{\cycidx} \gamma^2_{\cycidx},
    % \beta_{\alpha_{\cycidx}} = \frac{|\hat{S}_{xz}(\omega_k, \alpha_{\cycidx})|^2}{\sum_{\gamma \in \CyclicSet} |\hat{S}_{xz}(\omega_k, \gamma)|^2}.
$
where ${\gamma^2_{\cycidx} = |\hat{S}_{XZ}(\omega_k, \alpha_{\cycidx})|^2 / (\hat{S}_X(\omega_k)\,\hat{S}_Z(\alpha_p))}$ is the squared cyclic coherence between the noisy input and the noisy output.
% \end{align}
% This choice of the weights aims to give more relevance to elements that show high spectral correlation.
% In practice, the fundamental frequency and the cyclic set need to be estimated.
\section{Experiments}\label{sec:experiments}
The experiments compares the Wiener estimator of \cref{eq:wie_est}, based on a WSS model, and the estimator of \cref{eq:h_alpha_est}, which relies on the CS model, on the system identification task.

A limitation of the cyclic estimator is that it requires knowledge of the set of cyclic frequencies $\CyclicSet$.
In this work, the fundamental frequency of speech $\hat{f}_0(\ell)~[\si{\hertz}]$ that determines the set of cycles $\CyclicSet$ is estimated from the clean input $\fourier{S}_k(\ell)$ using the PYIN algorithm \cite{mauch_pyin_2014}
% \footnote{\url{librosa.org/doc/0.10.1/generated/librosa.pyin}} 
and converted to normalized angular frequency as $\hat{\omega}_0(\ell) = (2 \pi / f_s)\hat{f}_0(\ell)$.
The fundamental frequency $\hat{\omega}_0(\ell)$ may lie in the range $\mu_0 \pm \sigma_0$, where $\mu_0$ is the mean of $\hat{\omega}_0(\ell)$ over time and $\sigma_0$ is the standard deviation.
% $\mu_0 = L^{-1} \sum_{\ell = 0}^{L-1} \hat{\omega}_0(\ell)$ and 
% $\sigma_0 = \std{(\hat{\omega}_0(\ell))}$.
% \begin{align}
%     % \mu_0 = L^{-1} \sum_{\ell = 0}^{L-1} \hat{\omega}_0(\ell), \quad \sigma_0 = \std{(\hat{\omega}_0(\ell))}.
%     \mu_0 = \frac{1}{L} \sum_{\ell = 0}^{L-1} \hat{\omega}_0(\ell),~\sigma_0 = \sqrt{\frac{1}{L} \sum_{\ell = 0}^{L-1}(\hat{\omega}_0(\ell) - \mu_0)^2} 
%     .
% \end{align}
The estimated set $\hat{\CyclicSet}$ contains the cyclic frequencies $\alpha_p$ which are multiple of the fundamental:
$
% \begin{align}\label{eq:set_harmonics_cyclic}
\hat{\CyclicSet} = 
\{\alpha_p:~ 
(\mu_0 - \sigma_0)h
\leq \alpha_p \leq 
(\mu_0 + \sigma_0)h
\},
% \end{align}
$
where $h = 1, \ldots, H$.
As the harmonics of speech are more discernible for lower frequencies, we choose $H$ such that only frequencies smaller than $\SI{4}{\kilo\hertz}$ are considered.
% The set $\mathcal{I}_k$ of spectral frequencies that are multiple of the fundamental frequency follows from \cref{eq:set_harmonics_cyclic} by replacing the cyclic resolution $\Del \alpha$ by the spectral resolution $\Del f$.
% If instead of indices we collect spectral frequencies, we use $\SpectralSet$.

% \subsection{Simulation results}
\begin{figure}[tb]
\vspace{-1em}
    \centering
    \begin{subfigure}[t]{0.99\linewidth}
        \centering
        \hspace*{-0.4cm}
        \includegraphics[width=\linewidth]{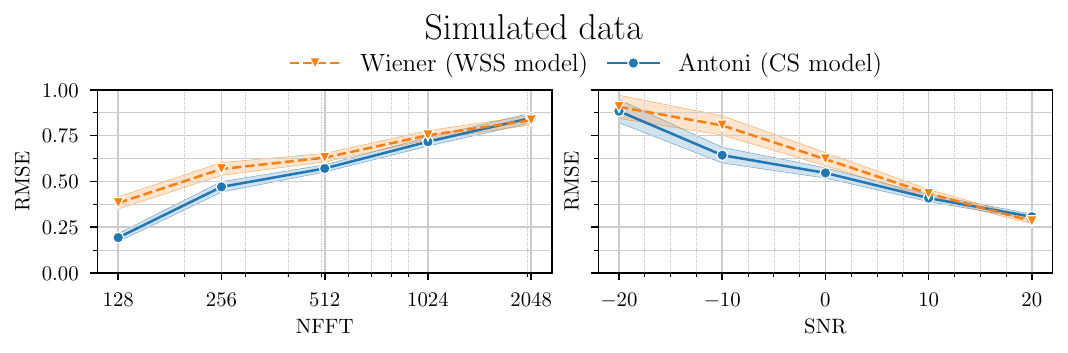}
        \label{fig:sub_errors_sim}
    \end{subfigure}
    \begin{subfigure}[t]{0.99\linewidth}
    \vspace{-2em}
        \centering
        \hspace*{-0.4cm}
        \includegraphics[width=\linewidth]{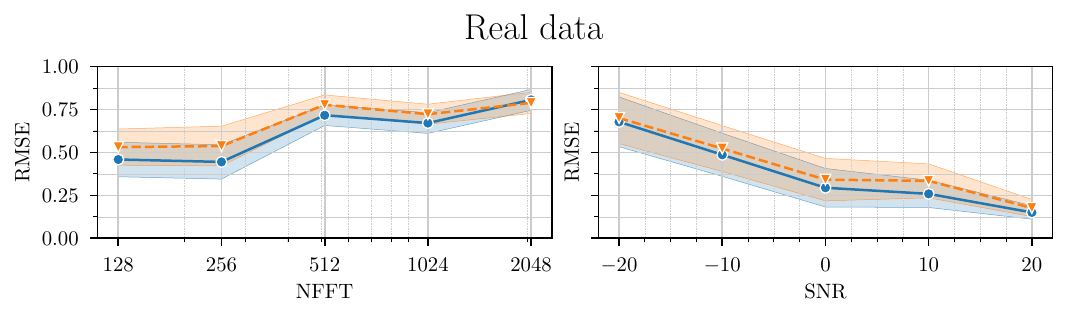}
        \label{fig:sub_errors_real}
    \end{subfigure}
    \vspace*{-0.em} % reduce space between caption and following text
    \caption{Root-mean-squared error between estimated and actual transfer function $a_k$ for both simulated (top) and real voiced speech (bottom).}
    \label{fig:errors}
    \vspace*{-1.5em} % reduce space between caption and following text
\end{figure}
% np.exp(-10 * np.arange(num_samples_h) / num_samples_h)
% np.sqrt(np.sum(np.abs(h) ** 2))
The experiments in \cref{fig:errors} compare the system identification performance of the classic Wiener estimator $\hat{a}^{\text{Wie}}$ and the cyclic estimator $\hat{a}^{\text{Ant}}$ as a function of the DFT length $K$ (left column) and input SNR (right column).
An LTI system $a(n)$ is simulated in the time-domain by drawing $K$ iid samples from $\mathcal{U}(-1,1)$ and applying a window $w_d(n) = e^{-10n/K}$. The system is then normalized to unitary energy as $a(n) \leftarrow a(n) ({\sum_n a(n)^2})^{-\frac{1}{2}}$.
The LTI system is either excited by voiced speech recordings from the UTD North Texas vowel database sampled at $f_s = \SI{16}{\kilo\hertz}$ \cite{assmann_time-varying_2000}, or by simulated signals that follow \cref{eq:sin_rnd_amp_SUM}.
The fundamental frequency of the simulated signals is randomly drawn from $\mathcal{U}(90, 250)\si{\hertz}$.
The cyclic spectra are estimated using the ACP estimator of \cref{eq:acp_estimator}.
The default value for the number of DFT points is $K=256$, and the default input SNR is $\SI{0}{\dB}$.
$w(n)$ is the Hann window, and the block-shift is set to $R = K / 3$ as suggested in \cite{antoni_cyclic_2007}.
The output noise is fixed at $\SI{40}{\dB}$ SNR.
Results are averaged over $40$ Montecarlo realizations with different noises, speech samples, and impulse responses.
Lines in the plot correspond to the mean values, while shaded areas represent the 95\% confidence intervals.
The performance metric is the root-mean-squared error (RMSE) between $a_k$ and $\hat{a}_k$, averaged over frequencies in $\hat{\CyclicSet}$.
For both simulated and real data, we observe that the cyclic estimator is equal to or better than the benchmark algorithm, especially for a smaller number $K$ of DFT points.
One reason for this is that smaller $K$ implies wider spectral bins and coarser spectral resolution $\Del \omega$, while the cyclic resolution $\Del \alpha$ is unchanged. Therefore, the ratio $\Del \omega / \Del \alpha$ increases, reducing the variance of the SCD estimate (\cref{ssec:est_spec_corr}).
While our results demonstrate promising performance under controlled conditions, it is important to note the practical challenges of estimating the fundamental frequency in noisy settings, which could limit applicability in dynamic environments.
\vspace{-1em}
\section{Conclusion}
The WSS model is inadequate for spectrally correlated processes such as speech.
This paper introduced a novel CS model for voiced speech that accounts for correlations across harmonic frequencies.
Experiments have shown that the new model can lead to improved system identification performance.
Moreover, the versatility of the proposed approach extends to various speech-related tasks, including PSD estimation, beamforming, and source separation, suggesting promising directions for future research.

\FloatBarrier
% \newpage
% \bibliographystyle{alpha}  % best for draft/debugging
\bibliographystyle{IEEEbib}
\bibliography{references}
\end{document}

%% file: definitions.tex
\usepackage{amsmath,graphicx}
\usepackage[breaklinks]{hyperref}
\usepackage{mathtools}
\usepackage{algorithmicx}
\usepackage{algorithm}
\usepackage{algpseudocode}
\usepackage{xcolor}
\usepackage{bm}
\usepackage{amssymb} % for the \shortminus
\usepackage{siunitx}
\usepackage{enumitem}
\usepackage[capitalise]{cleveref}
\usepackage{mathtools}
\usepackage{placeins}
\usepackage{booktabs}
\usepackage{microtype}
\usepackage{todonotes}
\usepackage{lipsum}
\usepackage{snaptodo}

\hypersetup{
    colorlinks,
    linkcolor={red!90!black},
    citecolor={blue!90!black},
    urlcolor={blue!90!black}
}

\DeclareMathSymbol{\shortminus}{\mathbin}{AMSa}{"39}

\newcommand{\Del}{\mathop{}\!\Delta}

\DeclarePairedDelimiter\brackets{[}{]}
\DeclarePairedDelimiter\ceil{\lceil}{\rceil}
\DeclarePairedDelimiter\floor{\lfloor}{\rfloor}

\DeclareMathOperator{\expectedValue}{\mathbb{E}}

\newcommand{\E}[1]{\expectedValue\brackets*{#1}}

\newcommand{\ie}{i.e., }

\long\def\hide#1{}

\newcommand{\red}[1]{\textcolor{red}{\textbf{#1}}}

\NewDocumentCommand{\grad}{e{_^}}{%
  \mathop{}\!% \mathop for good spacing before \nabla
  \mathop{}\mspace{-1mu}% \mathop for good spacing before \nabla
  \nabla
  \IfValueT{#1}{_{\!#1}}% tuck in the subscript
  \IfValueT{#2}{^{#2}}% possible superscript
  \mspace{-1mu}
}

\newcommand\scalemath[2]{\scalebox{#1}{\mbox{\ensuremath{\displaystyle #2}}}}
\newcommand\scalemathinline[2]{\scalebox{#1}{\mbox{\ensuremath{\textstyle #2}}}}

\let\ORGhypersetup\hypersetup
\protected\def\hypersetup{\ORGhypersetup}
\makeatletter
\def\th@plain{%
  \thm@notefont{}% same as heading font
  \itshape % body font
}
\def\th@definition{%
  \thm@notefont{}% same as heading font
  \normalfont % body font
}
\makeatother
\usepackage[font=footnotesize,position=bottom,belowskip=0pt,aboveskip=0pt]{caption}
\usepackage[font=footnotesize,position=bottom,belowskip=0pt,aboveskip=0pt]{subcaption}


\input{definitions_iwaenc}

%% file: definitions_iwaenc.tex
\def\funPer{P}
\def\CyclicSet{\mathcal{A}}
\def\cycidx{p}

\def\FourierPair{\overset{\scriptscriptstyle \mathcal{F}}{\longleftrightarrow}}
\def\xmod{X_N^{(\alpha_{\cycidx})}} % modulated in time domain
\def\sAmpSingleNameOnly{S_{\text{amp}}}
\def\sAmpSingle{\sAmpSingleNameOnly^{h}(n)}
\def\sPhaseSingle{S_{\text{ph}}^{h}(n)}

\def\autocorrAmpSingle{r_{\text{amp}}}

\def\sPhase{S_{\text{ph}}(n)}
\def\sAmp{S_{\text{amp}}(n)}
\def\sPhaseRealization{s_{\text{ph}}(n)}
\def\sAmpRealization{s_{\text{amp}}(n)}

\def\sReal{s_{\text{real}}(n)}
\def\numRealizationsPlot{200~}

\def\cycCorr{c_X(\alpha_{\cycidx}, \tau)} % R_x(\alpha_{\cycidx}, \tau)

\newcommand{\proc}[1]{\{{#1}\}}  % stands for "random process"
\newcommand{\fourier}[1]{\tilde{#1}}  % stands for "fourier transformed"
\newcommand{\smalleq}[1]{\scalemathinline{0.9}{#1}}  % stands for "small equation"